\documentclass[11pt]{article}

\usepackage{amsmath}
\usepackage{graphicx}
\usepackage[squaren]{SIunits}
\usepackage[unicode]{hyperref}
\usepackage{cite}

\newcommand{\gfrac}[2]{\displaystyle\frac{#1}{#2}}

\begin{document}

\title{Heteroscedasticity and angle resolution in high-energy particle tracking:
revisiting 
``Beyond the $\sqrt{\mathrm{N}}$ limit of the least squares resolution and the lucky model'', by 
G.~Landi and G.~E.~Landi}

\author{D.~Bernard, \\
LLR, Ecole Polytechnique, CNRS/IN2P3, 91128 Palaiseau, France}

\maketitle 

\begin{abstract}
 I re-examine a recent work by G. Landi and G. E. Landi.
 [arXiv:1808.06708 [physics.ins-det]], in which the authors claim
 that the resolution of a tracker can vary linearly with the number
 of detection layers, $N$, that is, faster than the commonly known
 $\sqrt{N}$ variation, for a tracker of fixed length, in case the
 precision of the position measurement is allowed to vary from layer
 to layer, i.e. heteroscedasticity, and an appropriate analysis method, a
weighted least squares fit, is used.
\end{abstract}

{\em keywords}:

Tracking,
weighted least squares,
homoscedasticity,
heteroscedasticity, 
Cramer-Rao Bound

\section{Introduction}

The momentum of charged particles, including the
magnitude and the direction, is one of the very basic observables on
which event reconstruction is built in particle physics.
In the case of a detector of fixed given length, $L$, containing a
number $N$ of detection layers, it is common wisdom that the precision
on the track angle improves as $1/\sqrt{N}$, asymptotically at large
$N$, so that the resolution improves as $\sqrt{N}$
(e.g. \cite{Regler:2008zza}, and references therein).

In a recent work, though, G. Landi and G. E. Landi. are claiming that
``{\sl A very simple Gaussian model is used to illustrate a new fitting result: a linear growth of the resolution with the number $N$ of detecting layers. This rule is well beyond the well-known rule proportional to $\sqrt{N}$ for the resolution of the usual fit}''
\cite{Landi:2018jdj} (and further developments in \cite{Landi:2019axo,Landi:2020mgd}).

As I didn't find the graphical pieces of evidence that were presented
in \cite{Landi:2018jdj} to support the allegation quite convincing, I
am trying here to re-examine the matter.
As in \cite{Landi:2018jdj}, I consider a simple situation of a tracker
consisting of equally-spaced parallel layers, without magnetic field,
and for which multiple scattering can be neglected.

\section{Weighted least squares straight-track fit }

Let's consider a tracker consisting of $N$ layers, $i = 1 \cdots N$, equally
spaced at position $x_i$ along the $x$ axis with spacing $D$.
Each detector $i$ measures the position in the transverse direction $y$ of each track traversing it, $y_i$,
with a Gaussian point spread function (PSF) with RMS $\sigma_i$.
We aim at fitting straight tracks
\begin{equation}
y = a x + b
\end{equation}
where $a$ and $b$ are the slope and the intercept of the track.
Minimization of the $\chi^2$,
\begin{equation}
 \chi^2 = \sum_{i=1}^{N} \left(\gfrac{y_i - (a x_i + b)}{\sigma_i}\right)^2
 ,
\end{equation}
provides the values of $a$ and $b$:
\begin{equation}
a = \gfrac{ s_{xy} s - s_x s_y }{ s_{x^2} s - (s_x)^2 }
 \quad
 \text{and}
 \quad
b = \gfrac{ s_y s_{x^2} - s_{xy} s_x }{ s_{x^2} s - (s_x)^2 }
\end{equation}
with precisions
\begin{equation}
\sigma_a =\sqrt{ \gfrac{ s}{ s_{x^2} s - (s_x)^2 } }
 \quad
 \text{and}
 \quad
\sigma_b =\sqrt{ \gfrac{ s_{x^2} }{ s_{x^2} s - (s_x)^2 } }
\end{equation}
and with
\begin{equation}
 s = \sum_{i=1}^{N} \gfrac{1}{\sigma_i^2} , 
 \quad
 s_x = \sum_{i=1}^{N} \gfrac{x_i}{\sigma_i^2} , 
 \quad
 s_y = \sum_{i=1}^{N} \gfrac{y_i}{\sigma_i^2} , 
 \quad
 s_{xy} = \sum_{i=1}^{N} \gfrac{x_i y_i}{\sigma_i^2} 
 \quad
 \text{and}
 \quad
 s_{x^2} = \sum_{i=1}^{N} \gfrac{x_i^2}{\sigma_i^2} .
\end{equation}
Given that $x_i = i D$, we have
\begin{equation}
 \sigma_a^2 =
 \gfrac{1}{D^2}
 \gfrac{ \sum_{i=1}^{N} \gfrac{1}{\sigma_i^2} } { \left( \sum_{i=1}^{N} \gfrac{i^2}{\sigma_i^2} \right) \left( \sum_{i=1}^{N} \gfrac{1}{\sigma_i^2} \right) - \left( \sum_{i=1}^{N} \gfrac{i}{\sigma_i^2} \right)^2 }
\label{eq:var:a}
\end{equation}

\section{Homoscedastic trackers}

In  case the precisions $\sigma_i$ are the same for all 
layers (homoscedasticity) and equal to a common value $\sigma$, 
eq. (\ref{eq:var:a}) simplifies and 
the precision of the measurement of the track angles boils down to 
\cite{Regler:2008zza}

\begin{figure}[t]
 \begin{center}
 \includegraphics[width=0.44\linewidth]{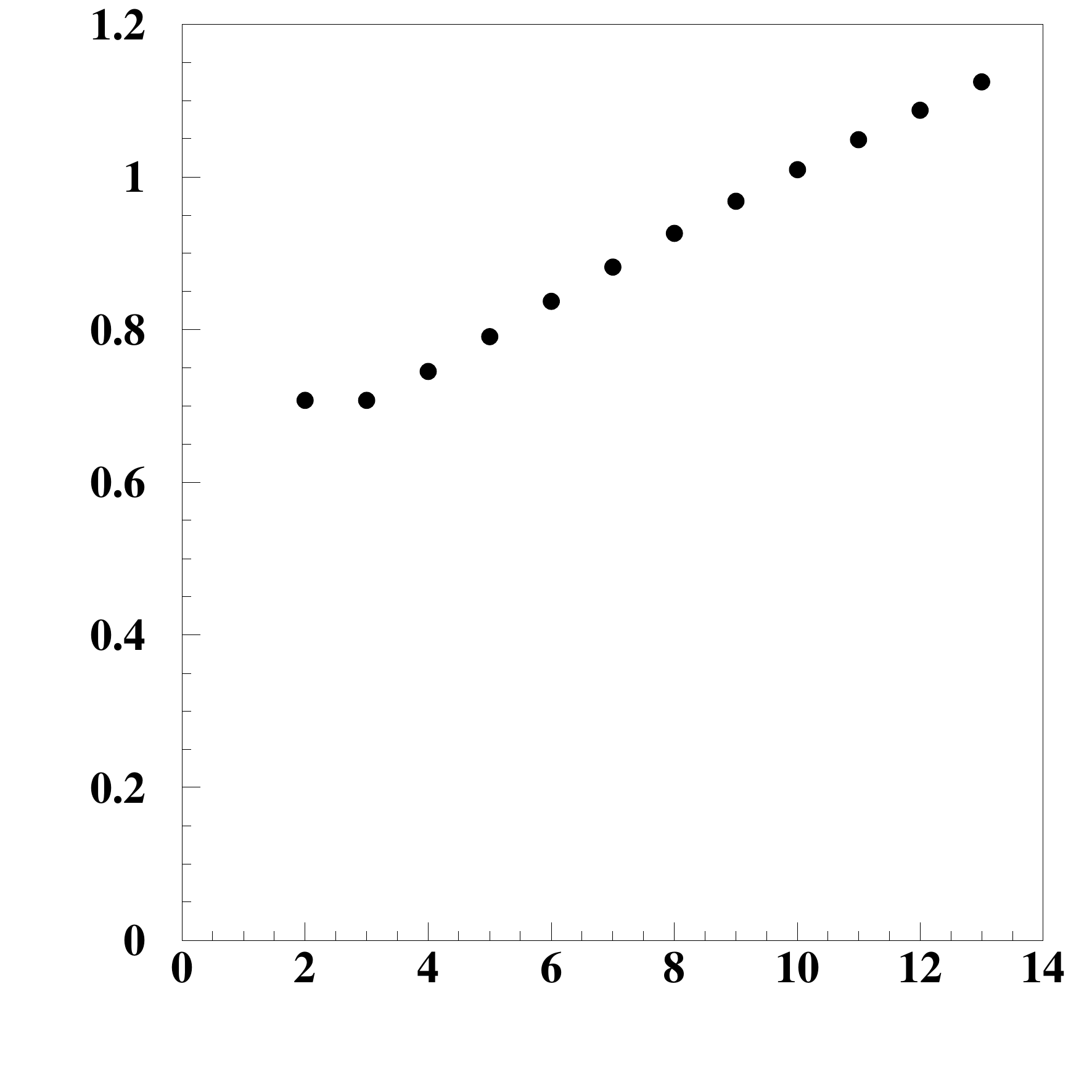}
 \put(-60,10){$N$}
 \includegraphics[width=0.44\linewidth]{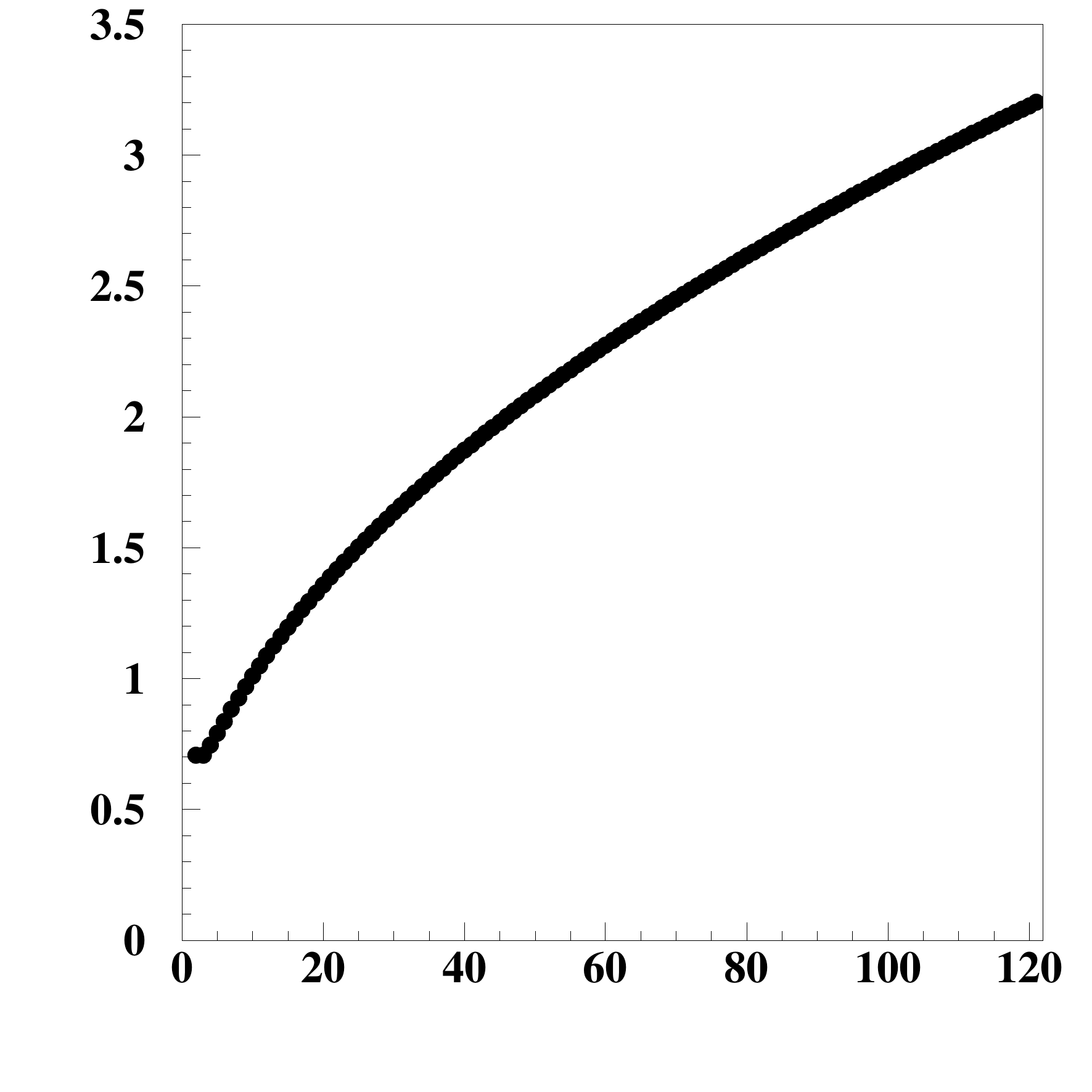}
 \put(-60,10){$N$}
 
 \caption{Homoscedastic trackers: single-Gaussian-distributed measurement precision: variation of the inverse precision,
 $1/ \sigma_a $, as a function of the number of detectors,
 for a fixed total detector length $L$, and for $\sigma / L = 1$.
 Left plot: up to $N-1=12$.
 Right plot: up to $N-1=120$.
\label{fig}
}
 \end{center}
\end{figure}

\begin{equation}
 \sigma_a =
 \gfrac{2 \sigma }{L} \sqrt{\gfrac{3 (N-1)}{N(N+1)}}
\quad =\quad 
 \gfrac{2 \sigma }{D} \sqrt{\gfrac{3 }{(N-1)N(N+1)}}
 , 
\end{equation}

where $L = (N-1) D$ is the total length of the detector.
For $N=1$, $ \sigma_a $ is undefined as was expected for an angle
measurement.
For trackers with a large number of detection layers, and for a total
length $L$ being kept constant, the precision of the measurement of
the angle varies asymptotically as $\sqrt{12 / N} \, \sigma / L$, and
the resolution as $\sqrt{N / 12} \, L / \sigma$.

It seems clear from the variation of the inverse precision,
$1/ \sigma_a $, as a function of the number of detectors,
Fig. \ref{fig}, that the linear variation with $N$ alluded in
\cite{Landi:2018jdj} is an ``impression'' when focusing attention on
the very smallest numbers of detectors (left plot), while the
asymptotic $\sqrt{N}$ variation is clearly visible for larger numbers
(right plot).

\section{Heteroscedastic trackers}

I now turn to the two-Gaussian toy model that G.~Landi and G.~E.~Landi
\cite{Landi:2018jdj} have used as being a good approximation of
tracking with silicon strip detectors, and with which they say they
observe a linear growth.
The point spread function consists of two Gaussians with different
standard deviations,
the first one with $\sigma_1 = 0.18$ with a probability of 80\,\% and the
second one with $\sigma_2 = 0.018$ with a probability of 20\,\%
\cite{Landi:2018jdj}.

With this model, fitting each track with a weighted least squares
provides values of $a$ with a Gaussian probability density function
with standard deviation $\sigma_a$ given by eq. (\ref{eq:var:a}),
(as demonstrated by the pull distribution, that is found to follow a perfect
${\cal N}(0,1)$ distribution) 
but the value of $\sigma_a$ varies from track to track, depending on
the distributions of the precisions of the measurements
($\sigma_1$ or $\sigma_2$) along the track.
As the $a$ distribution of the whole event sample is not
Gaussian-distributed, I use the same method as in
\cite{Landi:2018jdj} to obtain a samplewise estimate of $1 / \sigma_a$, that is,
the maximum of the $a$ distribution.

\begin{figure}[tbh]
 \begin{center}
 \includegraphics[width=0.44\linewidth]{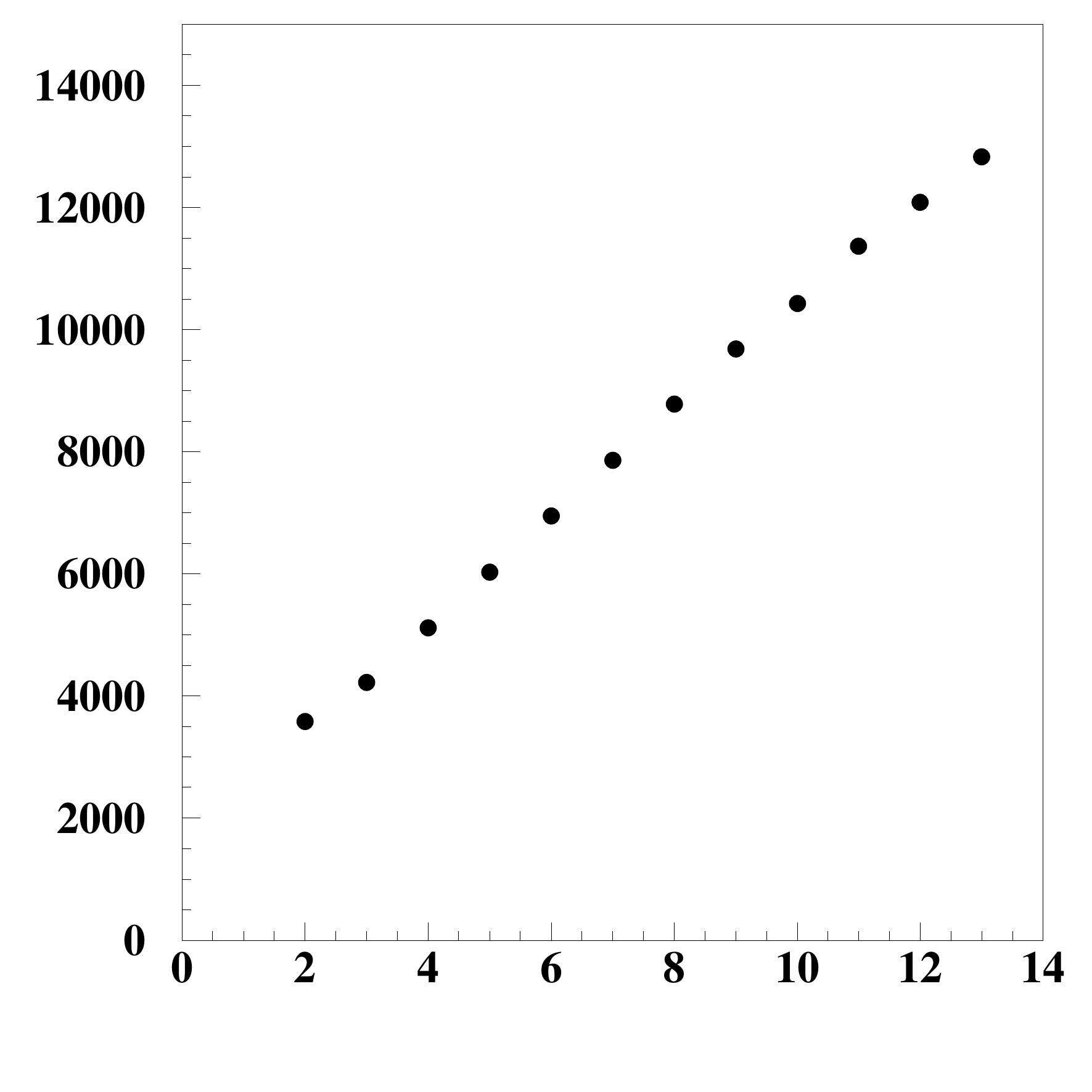}
 \put(-60,10){$N$}
 \includegraphics[width=0.44\linewidth]{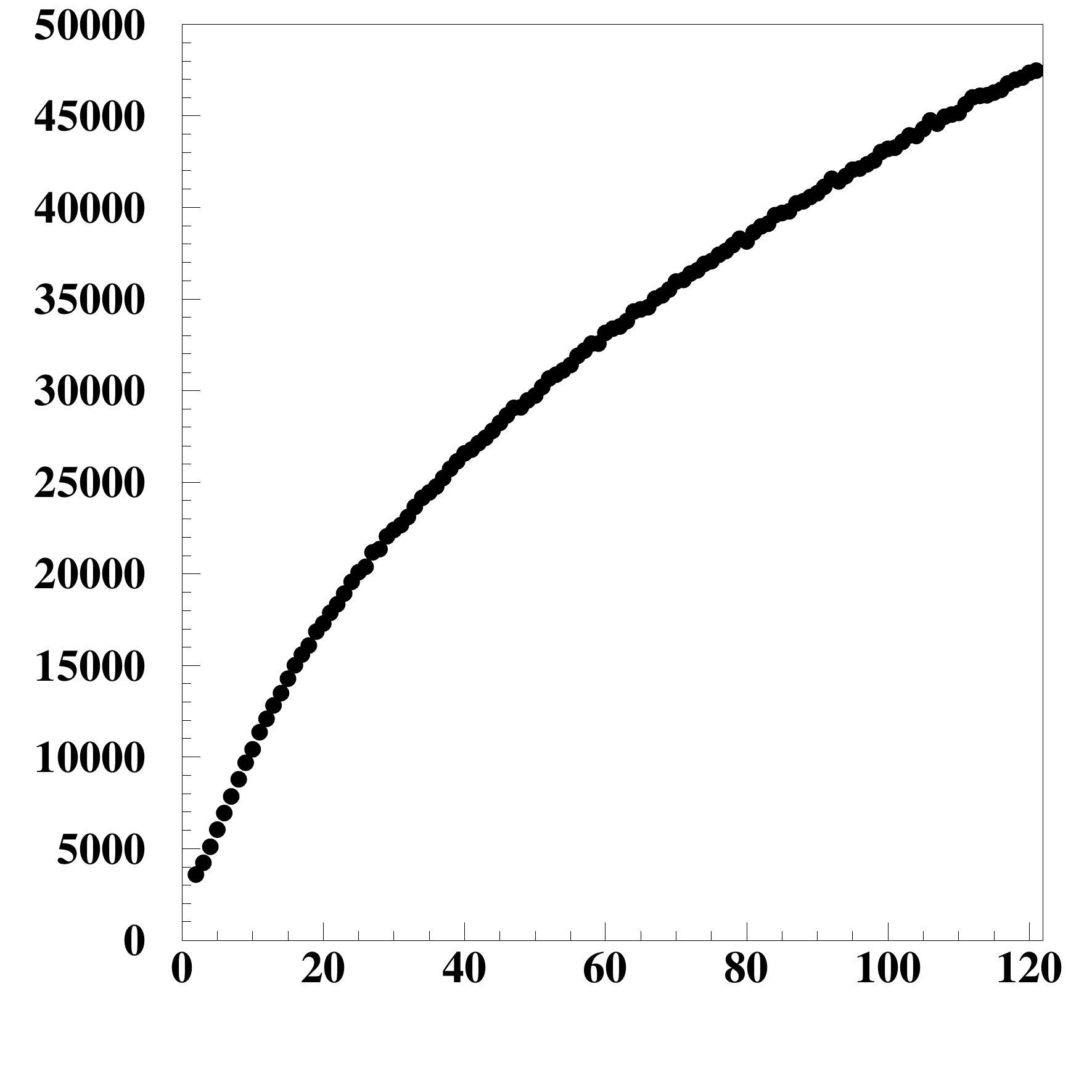}
 \put(-60,10){$N$}

\caption{Heteroscedastic trackers: double-Gaussian-distributed
 measurement precision: variation of the height of the $a$ peak at
 maximum, as a function of the number of detectors, for a fixed total
 detector length $L$. 
Left plot: up to $N-1=12$. Right plot: up to $N-1=120$.
\label{fig:2}
}
 \end{center}
\end{figure}

\begin{figure}[tbh]
 \begin{center}
 \includegraphics[width=0.44\linewidth]{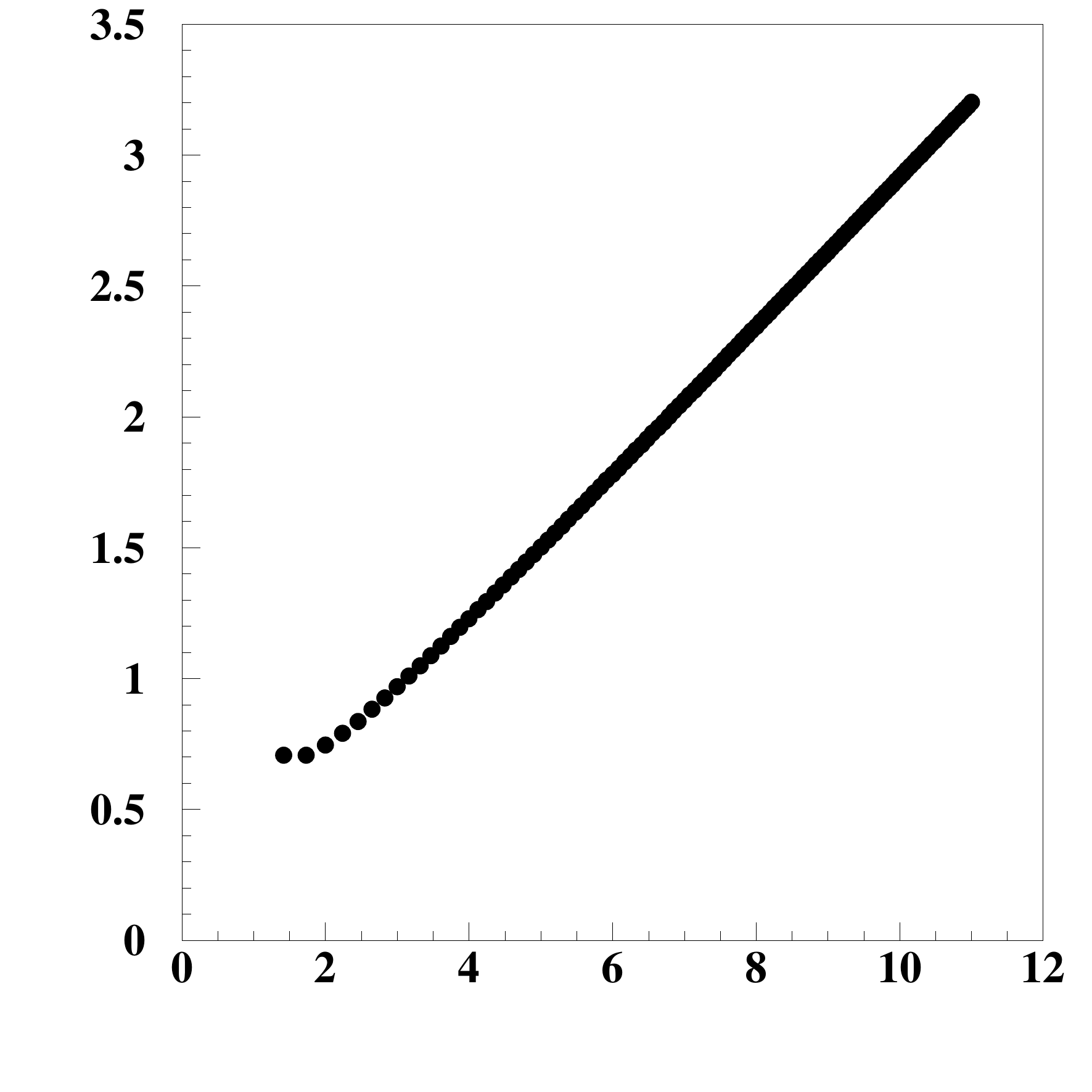}
 \put(-60,10){$\sqrt{N}$}
 \includegraphics[width=0.44\linewidth]{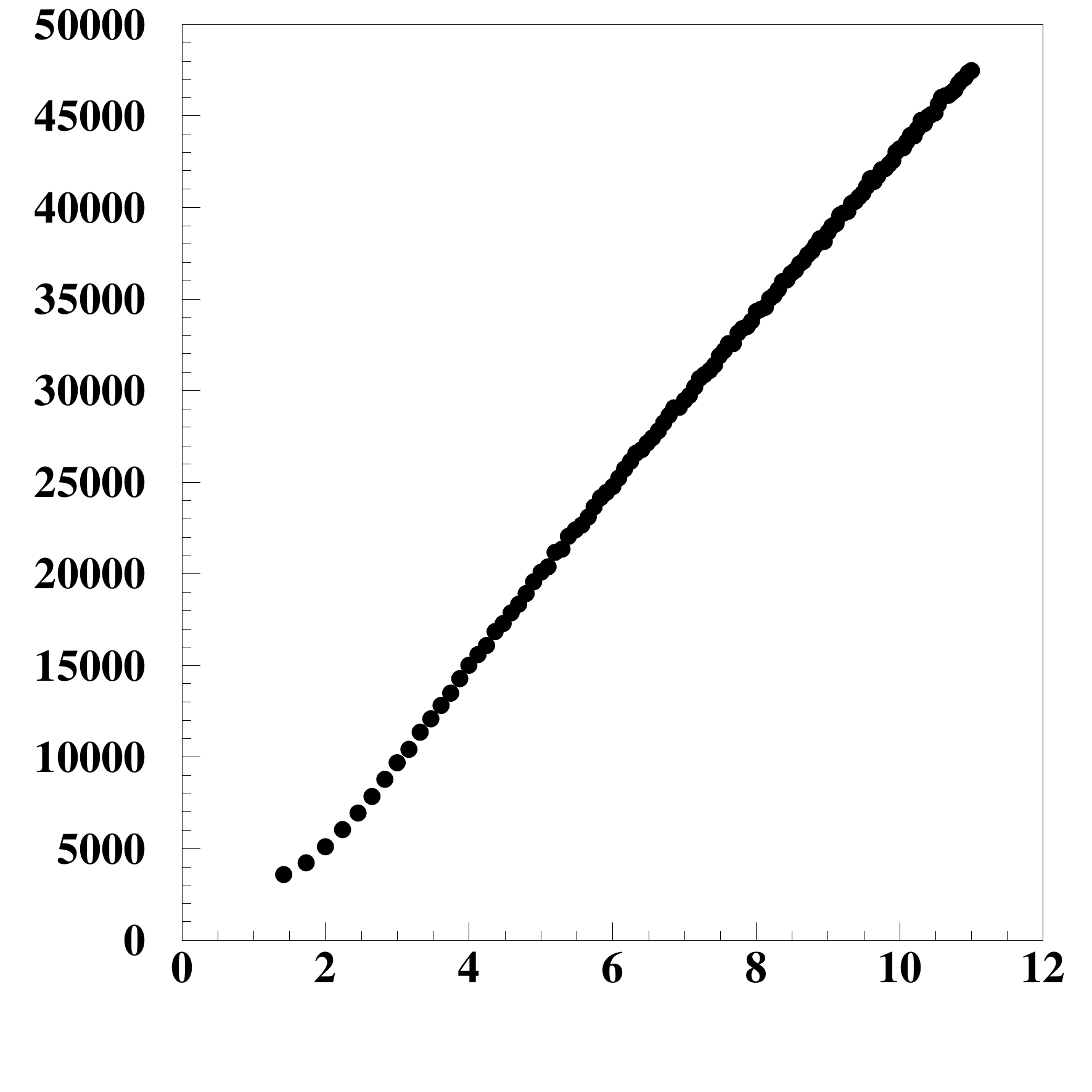}
 \put(-60,10){$\sqrt{N}$}
 \caption{Variation with $\sqrt{N}$ of the estimators shown in Figs.
  \ref{fig} and \ref{fig:2}, up to $N-1=120$.
Left: Homoscedastic trackers: single-Gaussian-distributed measurement
precision: variation of the inverse precision, $1/ \sigma_a $.
Right: Heteroscedastic trackers: double-Gaussian-distributed
measurement precision: variation of the height of the $a$ peak at
maximum.
\label{fig:sq}
}
 \end{center}
\end{figure}

For a small number of measurements, I do seem to observe a linear
growth (Fig. \ref{fig:2} left), as claimed in \cite{Landi:2018jdj},
but also in the same way as for the homoscedastic single-Gaussian
measurement examined in the previous section.
At large values of $N$, I obtain a $\sqrt{N}$-like variation,
something which is more easily observed on the variation with $\sqrt{N}$
(Fig. \ref{fig:sq}).

\section{Conclusion}

The present work does confirm that for tracking detectors consisting
of a small number of layers, the angle resolution seems to vary as $N$
as was shown in Fig. 2 of \cite{Landi:2018jdj}.
Examination of detectors consisting of a large number of layers,
though, shows a variation of the resolution as $\sqrt{N}$, compatible
with common wisdom.
Neither homoscedasticity nor heteroscedasticity are found to play any
role in the matter, in contrast with what alleged in
\cite{Landi:2018jdj}.

% \clearpage


\begin{thebibliography}{00} \small
 

%\cite{Regler:2008zza}
\bibitem{Regler:2008zza}
 M.~Regler and R.~Fruhwirth,
 ``Generalization of the Gluckstern formulas. I: Higher orders, alternatives and exact results,''
\href{https://inspirehep.net/literature/790377}{Nucl.\ Instrum.\ Meth.\ A {\bf 589} (2008) 109}.
% doi:10.1016/j.nima.2008.02.016


%\cite{Landi:2018jdj}
\bibitem{Landi:2018jdj}
G.~Landi and G.~E.~Landi,
``Beyond the $\sqrt{\mathrm{N}}$ limit of the least squares resolution and the lucky model,''
\href{https://arxiv.org/abs/1808.06708}{[arXiv:1808.06708 [physics.ins-det]]}.


%\cite{Landi:2019axo}
\bibitem{Landi:2019axo}
G.~Landi and G.~E.~Landi,
``The Cramer-Rao Inequality to Improve the Resolution of the Least-Squares Method in Track Fitting,''
\href{https://inspirehep.net/literature/1762437}{Instruments \textbf{4} (2020) 2}
% doi:10.3390/instruments4010002
% [arXiv:1910.14494 [physics.ins-det]].

%\cite{Landi:2020mgd}
\bibitem{Landi:2020mgd}
G.~Landi and G.~E.~Landi,
``Proofs of non-optimality of the standard least-squares method for track reconstructions,''
\href{https://arxiv.org/abs/2003.10021}{[arXiv:2003.10021 [math.ST]]}.

\end{thebibliography}
\end{document}